\documentclass[reprint,superscriptaddress,amsmath,amssymb,prb]{revtex4-1}
\usepackage[dvipdfmx]{graphicx}% Include figure files
\usepackage{dcolumn}% Align table columns on decimal point
\usepackage{bm}
\usepackage{color}
\usepackage{amssymb}
\usepackage{comment}

\def\bmE{{\bm E}}
\def\bmA{{\bm A}}

\def\bmK{{\bm K}}
\def\bmP{{\bm P}}

\def\bmI{{\bm I}}
\def\bmk{{\bm k}}
\def\bmr{{\bm r}}

\def\bmp{{\bm p}}

\def\rmp{{\rm probe}} %probe
\def\rmP{{\rm pump}} %pump
\def\rmI{{\rm I}}
\def\rmH{{\rm H}}

\begin{document}

\title{Symmetry properties of attosecond transient absorption spectroscopy in crystalline dielectrics}

\author{Shunsuke Yamada}
\affiliation{Center for Computational Sciences, University of Tsukuba, Tsukuba 305-8577, Japan}
\author{Kazuhiro Yabana}
\affiliation{Center for Computational Sciences, University of Tsukuba, Tsukuba 305-8577, Japan}

\date{\today}

\begin{abstract}
We theoretically investigate a relation between the crystalline symmetry and the transient modulation 
of optical properties of crystalline dielectrics in pump-probe measurements using intense pump
and attosecond probe fields.
When the photon energy of the pump field is much below the bandgap energy, the modulation of the optical conductivity
is caused by the intraband electronic motion, that is, the dynamical Franz-Keldysh effect.
We analytically investigate symmetry properties of the modulated optical conductivity utilizing the Houston function,
and derive a formula that relates the temporal oscillation in the absorption with the transformation properties of 
the modulated optical conductivity.
To verify the validity of the formula, we perform real-time first-principles calculations based on the 
time-dependent density functional theory for a pump-probe process taking 4H-SiC crystal as an example. 
\end{abstract}
\maketitle

\section{Introduction}

Recent progress in attosecond metrology has made it possible to investigate electron motion
in solids in a time-scale less than a cycle of an optical pulse\cite{Goulielmakis2007}.
To explore the ultrafast electron motion, the attosecond transient absorption spectroscopy (ATAS) method
has been often utilized\cite{Goulielmakis2010,Wang2010,Holler2011}. 
In the ATAS measurements, a strong pump pulse of visible or infrared frequency
and a weak attosecond probe pulse irradiate on a thin film with a certain time delay.
The modulation of the optical absorption of the probe pulse is used to explore the ultrafast change of
optical properties of the thin film.
In a number of measurements in dielectric materials, 
field-driven oscillations of the absorption change as a function of the pump-probe delay time have been observed
\cite{Schultze2013,Schultze2014,Mashiko2016,Lucchini2016,Moulet2017,Schlaepfer2018}. % 2Ωだけではない。

Depending on the choices of parameters of applied laser pulses and materials, 
several mechanisms that contribute to modulations in ATAS have been proposed.
Among them, the intraband motion of electrons induced by the strong pump electric field that causes
the dynamical Franz-Keldysh effect (DFKE) is considered to be one of the primary mechanisms.
The Franz-Keldysh effect (FKE) is a modulation of the absorption properties of dielectrics at around 
the bandgap energy under a static electric field and has been extensively investigated since it was first discussed 
more than half a century ago
\cite{Franz1958,Keldysh1958,Tharmalingam1963,Seraphin1965,Nahory1968,Shen1995,Wahlstrand2010,Duque-Gomez2015}.
The DFKE is the modulation under an alternating electric field and has also been investigated in a
number of literatures
\cite{Yacoby1968,Jauho1996,Nordstrom1998,Srivastava2004,Mizumoto2006,Ghimire2011,Chin2000}.

In Ref.~\onlinecite{Otobe2016}, it has been analytically shown that the frequency of 
ATAS modulation is twice the pump frequency $\Omega$ in an isotropic two-band model. 
It was also confirmed numerically by the first-principles calculations conducted for a crystalline diamond.
There are some cases where oscillations different from $2\Omega$ frequency:
an oscillation with frequency $\Omega$ was theoretically described in two-dimensional material \cite{Sato2018},
while an oscillation with $3\Omega$ has been reported in GaN \cite{Mashiko2016}.
Although a relation between the crystalline symmetry and the static FKE has been discussed
\cite{Wahlstrand2010,Wahlstrand2011,Wahlstrand2011-2},
such analysis has not been reported for the time-resolved DFKE.
We note that, for the high harmonic generation in solids, dynamical symmetries that involve temporal as well as
spatial invariances have been extensively studied \cite{Tang1971,Alon1998,Averbukh1999,Simon1968,Konishi2014,Chen2014,McPherson1987,Krausz2009,Ceccherini2001,Averbukh2001,Alon2002,Liu2016,Saito2017,Neufeld2019}.
%\cite{Tang1971,Alon1998,Averbukh1999,Simon1968,Konishi2014,Chen2014,McPherson1987,Krausz2009,Eichmann1995,Milosevic2000,Milosevic2000a,Averbukh2002,Fleischer2014,Kfir2015,Milosevic2015,Mauger2016,Bandrauk2016,Ceccherini2001,Averbukh2001,Alon2002,Liu2016,Saito2017,Neufeld2019}.

In the present paper, we first theoretically investigate a relation between symmetry properties of a dielectric crystal 
and its DFKE response using an analytical framework extending that introduced in Ref.~\onlinecite{Otobe2016}.
We will show that the oscillation property of the time-resolved DFKE response is determined from the transformation properties 
of the pump-modulated conductivity tensor under the point group symmetry operation of the crystal.
We next numerically verify this statement by performing first-principles calculations employing the time-dependent 
density functional theory (TDDFT)\cite{Runge1984} for hexagonal silicon carbide (4H-SiC) crystal.
We solve the time-dependent Kohn-Sham (TDKS) equation in the time domain and directly simulate 
ATAS in the crystal. We show that the calculated results under several polarization conditions 
confirm the validity of our analytical results.

This paper is organized as follows: 
In Sec.~\ref{sec:absorption}, a general description of the ATAS is considered introducing
a frequency-resolved absorbance function.
Section~\ref{sec:analytic} describes analytical considerations for DFKE and its symmetry dependence.
In Sec.~\ref{sec:first-principles}, we present a formalism and results of real-time first-principles calculations.
In Sec.~\ref{sec:conclusion}, a conclusion is presented.
%Finally, an appendix describes the perturbation expansion of the time-resolved conductivity with respect to the pump field.

\section{Transient absorption in a unit cell of solids \label{sec:absorption}}

We first provide a theoretical description of ATAS extending and clarifying the description presented in Ref.~\onlinecite{Otobe2016}.
We introduce a transient conductivity that depends both on the time and frequency.
In the derivation, we put emphasis on the relation between the frequency-dependent absorption
spectrum of the probe pulse and the transient conductivity under the intense pump field.

\subsection{Frequency-dependent modulation of probe absorption}

We consider a unit cell of dielectrics under irradiation of intense pump and weak probe fields.
In ATAS, the frequency of the probe  pulse is much higher than that of the pump pulse.
We assume that there is no overlap between two pulses in the frequency domain.
Assuming that both wavelengths of the pump and the probe pulses are sufficiently longer than
the size of the unit cell, we employ a dipole approximation treating electric fields of the pump 
and the probe pulses as spatially-uniform fields. This assumption will be justified for attosecond
probe fields up to, at least, a few tens of eV.

We express the electric field applied to the unit cell as $\bmE(t)$ .
It induces electron dynamics and charge current in the unit cell.
We express the macroscopic charge-current density as $\bmI(t)$ that is obtained as a spatial 
average of the microscopic current density in the unit cell.
The optical absorption can be evaluated from the work done by the electric field $\bmE(t)$
to electrons in the unit cell. It  is expressed as
\begin{equation}
W[\bmE] = \int dt \bmI[\bmE](t) \cdot \bmE(t).
\label{eq:work}
\end{equation}
In the above expression, we write the work as $W[\bmE]$ and the macroscopic current density as $\bmI[\bmE](t)$ to
stress that they are defined for the electric field $\bmE(t)$.

To discuss frequency-resolved signals, we rewrite the work as the integral over frequency,
\begin{equation}
W[\bmE] = \frac{1}{\pi} \int_0^\infty d\omega \frac{d \tilde W}{d\omega},
\end{equation}
where the frequency-resolved work is defined by
\begin{equation}
\frac{d\tilde W}{d\omega} = {\rm Re} \left[ \tilde \bmI[\bmE](\omega) \cdot \tilde \bmE^*(\omega) \right].
\end{equation}
Here we introduce time-frequency Fourier transformations, for example, by
\begin{equation}
\tilde \bmE(\omega) = \int_{-\infty}^{\infty} dt e^{i\omega t} \bmE(t).
\end{equation}

We consider three cases, pump only, probe only, and pump plus probe.
Expressing the pump field as $\bmE^\rmP(t)$ and the probe field as $\bmE^\rmp(t)$, we have
\begin{equation}
\frac{dW^\rmP}{d\omega} = {\rm Re} \left\{ \tilde \bmI[\bmE^\rmP](\omega) \cdot \tilde \bmE^{\rmP*}(\omega) \right\},
\end{equation}
\begin{equation}
\frac{dW^\rmp}{d\omega} = {\rm Re} \left\{ \tilde \bmI[\bmE^\rmp](\omega) \cdot \tilde \bmE^{\rmp*}(\omega) \right\},
\end{equation}
\begin{eqnarray}
&&\frac{dW^{\rmP+\rmp}}{d\omega} = \nonumber\\
&&  {\rm Re} \left\{ \tilde \bmI[\bmE^{\rmP+\rmp}](\omega) \cdot  \tilde \bmE^{\rmP + \rmp*} (\omega) \right\},
\label{WPp}
\end{eqnarray}
where $\tilde \bmE^{\rmP+\rmp}(\omega)$ is given by $\tilde \bmE^\rmP(\omega) + \tilde \bmE^\rmp(\omega)$.
The modulation in the ATAS is then given as the difference between $dW^{\rmP+\rmp}/d\omega$
and $dW^\rmp/d\omega$.

In the following development, we will consider a quantity $A(\omega)$ defined below instead of $dW^{\rmP+\rmp}/d\omega$,
\begin{equation}
A(\omega) =  {\rm Re} \left[ \delta \tilde I(\omega) \cdot \tilde \bmE^{\rmp,*} (\omega) \right],
\label{Aw}
\end{equation}
where we introduce a modulated current density of the probe pulse,
\begin{equation}
\delta \tilde{\bmI}(\omega) = \tilde \bmI[ \bmE^{\rmP+\rmp}](\omega) - \tilde \bmI[ \bmE^\rmP](\omega).
\label{dIomega}
\end{equation}
We note that $A(\omega)$ accurately describes the absorption in the frequency region of the probe pulse,
\begin{equation}
A(\omega) \simeq \frac{dW^{\rmP+\rmp}}{d\omega} \hspace{5mm} (\omega^\rmp_1 < \omega < \omega^\rmp_2 ),
\end{equation}
where $\omega^\rmp_{1/2}$ is the lower/upper bound of the frequency spectrum of the probe pulse.
To understand it, we note that $\tilde \bmE^\rmP(\omega)$ in $\tilde \bmE^{\rmP+\rmp}(\omega)$ of 
Eq.~(\ref{WPp}) does not contribute in the probe frequency region, since we assume that there is no
overlap between pump and probe fields in the frequency domain.
A term containing $\tilde \bmI[\bmE^\rmP](\omega)$ of Eq.~(\ref{dIomega}) in $A(\omega)$ does not contribute
in the probe frequency region, since we assume that frequencies of the pump and the probe pulses are
well separated. In practice, $\tilde \bmI[\bmE^\rmP](\omega)$ has a small contribution in the spectral region
of the probe pulse through nonlinear effects such as high harmonic generation. However, the  contribution will
be extremely small since it is related with high order nonlinear processes.
In the following, we will call $A(\omega)$ the frequency-resolved absorbance.
We note that $A(\omega)$ was treated as a central quantity to discuss the modulation of the dielectric property
in the literatures \cite{Otobe2016,Lucchini2016}.

\subsection{Conductivity under a strong pump field}

We consider the current difference introduced in Eq.~(\ref{dIomega}) in time domain,
\begin{equation}
\delta \bmI(t) = \bmI[\bmE^{\rmP+\rmp}](t) - \bmI[\bmE^\rmP](t).
\label{delta_current}
\end{equation}
Since we assume that the probe pulse is a weak perturbation, we may introduce a linear
constitutive relation,
\begin{equation}
\delta I_\alpha(t) = \sum_{\beta} \int dt' \sigma_{\alpha\beta}(t,t') E^\rmp_\beta(t'),
\label{def_conductivity}
\end{equation}
where $\alpha$ and $\beta$ denote the Cartesian components, $\alpha, \beta = x, y, z$.
Here we introduced a conductivity $\sigma_{\alpha\beta}(t,t')$ under the strong
pump field, $\bmE^\rmP(t)$. It depends on $t$ and $t'$ separately, since the pump field breaks the
translational invariance in time.

We may express the constitutive relation in the frequency representation as
\begin{equation}
\delta \tilde I_{\alpha}(\omega) = \sum_\beta \int d\omega' \tilde \sigma_{\alpha\beta}(\omega,\omega') \tilde E^\rmp_\beta(\omega'),
\end{equation}
where the frequency-dependent conductivity is defined by
\begin{equation}
\tilde \sigma_{\alpha\beta}(\omega,\omega') = \int \frac{dt dt'}{2\pi} e^{i\omega t - i\omega' t'} \sigma_{\alpha\beta}(t,t').
\label{sigma_ww}
\end{equation}
This conductivity is related to the ordinary conductivity in the absence of the
pump pulse, which we denote as $\tilde \sigma_{\alpha\beta}^0(\omega)$, by
\begin{equation}
\tilde \sigma_{\alpha\beta}[\bmE^\rmP=0](\omega, \omega') = \delta (\omega-\omega') \tilde \sigma_{\alpha\beta}^0(\omega).
\end{equation}

Using the conductivity thus defined, the frequency-resolved absorbance defined by Eq.~(\ref{Aw}) is expressed as
\begin{equation}
A(\omega) =
{\rm Re} \sum_{\alpha\beta} \int d\omega'
\tilde E^{\rmp*}_\alpha(\omega)
\tilde \sigma_{\alpha\beta}(\omega,\omega')
\tilde E^{\rmp}_\beta (\omega').
\end{equation}

\subsection{Periodic pump field}

We next assume that the pump field is periodic in time with a frequency $\Omega$,
\begin{equation}
\bmE^\rmP \left( t+ T_\Omega \right) = \bmE^\rmP(t),
\label{eq:periodicity_A}
\end{equation}
where $T_\Omega = 2\pi/\Omega$ is the period of the pump field.
We assume that the conductivity has the same periodicity.
\begin{equation}
\sigma_{\alpha\beta} \left( t+ T_\Omega,t'+T_\Omega \right) = \sigma_{\alpha\beta}(t,t').
\end{equation}
This assumption may be justified for systems without resonant absorption,
for example, a wide-gap dielectric under a pump pulse whose photon energy
is much below the bandgap energy.

To proceed, we write the two-time conductivity $\sigma_{\alpha\beta}(t,t')$ 
as $\sigma_{\alpha\beta}(t+s,t)$, and recognize it as
a function of $t$ and $s$. Then, it is a periodic function of $t$ with the period $T_\Omega$,
and we may introduce a Fourier decomposition,
\begin{equation}
\sigma_{\alpha\beta}(t+s,t) = \sum_{n=-\infty}^{\infty} e^{in\Omega t} \sigma_{\alpha\beta}^{(n)}(s),
\end{equation}
\begin{equation}
\sigma_{\alpha\beta}^{(n)}(s) = \frac{1}{T_\Omega} \int_0^{T_\Omega} dt e^{-in\Omega t} 
\sigma_{\alpha\beta}(t+s,t).
\end{equation}
Using this property, we may express $\tilde \sigma_{\alpha\beta}(\omega,\omega')$ defined by Eq.~(\ref{sigma_ww}) as
\begin{equation}
\tilde \sigma_{\alpha\beta}(\omega,\omega')
= \sum_n \delta (\omega - \omega' + n\Omega) \tilde \sigma_{\alpha\beta}^{(n)}(\omega),
\end{equation}
\begin{equation}
\tilde \sigma^{(n)}_{\alpha\beta}(\omega)
= \int ds e^{i\omega s} \sigma^{(n)}_{\alpha\beta}(s).
\end{equation}
Therefore, the conductivity $\tilde \sigma_{\alpha\beta}(\omega,\omega')$ contributes only when
$\omega-\omega'$ is equal to integer multiples of the pump frequency $\Omega$.
The frequency-resolved absorbance now becomes
\begin{equation}
A(\omega)= 
{\rm Re} \sum_{\alpha\beta n}
\tilde E^{\rmp*}_\alpha(\omega)
\tilde E^{\rmp}_\beta(\omega+n\Omega)
\tilde \sigma_{\alpha\beta}^{(n)}(\omega).
\end{equation}

\subsection{Transient absorption using impulsive probe pulse}

To proceed, we next specify the pulse shape of the probe pulse.
We first consider an impulsive field as an extreme case, 
\begin{equation}
E^\rmI_\alpha(t) = F_\alpha \delta(t-T),
\end{equation}
where $T$ specifies the time when the impulsive field is applied and 
$F_\alpha$ is a parameter that specifies the strength of the field in $\alpha$ direction.
For this impulsive field, the induced current is given by
\begin{equation}
\delta I^\rmI_\alpha(t) = \sum_\beta \sigma_{\alpha\beta}(t,T) F_\beta.
\end{equation}
Taking Fourier transformation of the both sides, we have
\begin{equation}
\delta \tilde I^\rmI_{\alpha}(\omega) = \sum_{\beta} \tilde \sigma^\rmI_{\alpha\beta}(T,\omega)
\tilde E^\rmI_\beta(\omega),
\end{equation}
where the conductivity $\tilde \sigma^\rmI_{\alpha\beta}(T,\omega)$ that depends on both time $T$
and frequency $\omega$ is introduced as
\begin{eqnarray}
\tilde \sigma^\rmI_{\alpha\beta}(T,\omega) 
&=&
\int dt e^{i\omega(t-T)} \sigma_{\alpha\beta}(t,T)
\nonumber\\
&=&
\sum_n e^{in\Omega T} \tilde \sigma_{\alpha\beta}^{(n)}(\omega).
\label{def_impulsive}
\end{eqnarray}
Using this conductivity, the frequency-resolved absorbance of Eq.~(\ref{Aw}) is expressed as
\begin{equation}
A(\omega) = {\rm Re}
\sum_{\alpha\beta} F_{\alpha}F_{\beta} \tilde \sigma_{\alpha\beta}^\rmI(T,\omega).
\end{equation}
In this way, we can introduce the conductivity $\tilde \sigma^\rmI_{\alpha\beta}(T,\omega)$ with mixed indices, 
the time $T$ that specifies when the impulsive field is applied and the frequency $\omega$ that specifies 
the frequency of the absorption of the probe pulse.

\subsection{Transient absorption using general probe pulse}

We next consider a more general case with a finite duration of the probe pulse.
For a probe pulse applied at time $t=T$, we express the time profile of the probe pulse by
\begin{equation}
E^\rmp_\alpha(t) = f_{\alpha}(t-T),
\end{equation}
where the function $f_{\alpha}(t)$ has a maximum at $t=0$.
The Fourier transform of the probe pulse is given by
\begin{equation}
\tilde E^\rmp_\alpha(\omega) = e^{i\omega T} \tilde f_\alpha(\omega).
\end{equation}
The phase of the function $\tilde f_\alpha(\omega)$ may not depend much
on the frequency.
For example, if $f_{\alpha}(t)$ is an even function of $t$, 
$\tilde f_\alpha(\omega)$ is a real function. If $f_{\alpha}(t)$ is an odd function of $t$, 
$\tilde f_\alpha(\omega)$ is a pure imaginary function. In both cases, the phase part of 
$\tilde f_\alpha(\omega)$ shows no frequency dependence.

The frequency-resolved absorbance of Eq.~(\ref{Aw}) is calculated as
\begin{equation}
A(\omega) = {\rm Re}
\sum_{\alpha\beta n} 
\tilde f^*_\alpha(\omega) \tilde f_\beta(\omega + n \Omega) 
e^{in\Omega T} \tilde \sigma^{(n)}_{\alpha\beta}(\omega).  
\label{abs_sigma}
\end{equation}
This result indicates again that the modulation in the absorption depends on the time $T$ 
through the frequency $\Omega$ and its multiples. 

In the ATAS, an extremely short attosecond pulse is used for the probe pulse.
For such pulses, we may approximate $\tilde f_\beta(\omega + n\Omega) \simeq \tilde f_\beta(\omega)$,
Then Equation (\ref{abs_sigma}) can be expressed as,
\begin{equation}
A(\omega)
\simeq
{\rm Re} \sum_{\alpha\beta}  
\tilde f^*_\alpha(\omega) \tilde f_\beta(\omega) \tilde \sigma_{\alpha\beta}^\rmI(T, \omega).
\end{equation}
We thus find that the frequency-resolved modulation in the absorption of the probe pulse can be
described using the impulsive time-resolved conductivity, $\tilde \sigma^\rmI_{\alpha\beta}(T,\omega)$,
if the probe pulse is sufficiently short.
In the next section, we focus on the symmetry properties of this conductivity.

\section{Analytic consideration \label{sec:analytic}}

\subsection{Conductivity using Houston function}

To investigate symmetry properties of the conductivity in the presence of a strong pump field, 
we utilize a model description in which the electronic system in a unit cell of the crystal  is described 
by a single-electron Bloch equation. We first consider a static problem,
\begin{equation}
{H}_{\bmk}u_{n{\bmk}}({\bmr}) \equiv
\left[ \frac{1}{2}({\bmp}+{\bmk})^2+V({\bmr}) \right]u_{n{\bmk}}({\bmr})=
\varepsilon_{n{\bmk}} u_{n{\bmk}}({\bmr}), 
\label{eq:gse}
\end{equation}
where ${H}_{\bmk}$ and $u_{n{\bmk}}$ are the effective single-electron Hamiltonian and the Bloch orbital in the ground state, respectively.
We next consider electron dynamics under a spatially-uniform electric field which is described by the vector potential ${\bmA}(t)$.
The Bloch orbital that describes electron dynamics under the electric field, $v_{n{\bmk}}({\bmr},t)$, follows the time-dependent
Schr\"odinger equation,
\begin{equation}
    i\frac{\partial}{\partial t}v_{n{\bmk}}({\bmr},t)=H_{\bmk+\frac{1}{c}\bmA(t)}\, v_{n{\bmk}}({\bmr},t).
    \label{eq:tdse}
\end{equation}
Here we assume that the same periodic potential $V({\bmr})$ is used in Eqs.~(\ref{eq:gse}) and (\ref{eq:tdse}).
The electric current density averaged over the unit cell is given by
\begin{eqnarray}
    {\bmI}(t) &=&-\frac{1}{V}\sum_{n{\bmk}} f_{n{\bmk}}
    \int_{\rm cell} d\bmr \,
    v^*_{n{\bmk}}(\bmr, t)  \nonumber\\ 
    &&\times \left( {\bmp} +\bmk + \frac{\bmA(t)}{c} \right) v_{n{\bmk}}(\bmr, t),
    \label{eq:I}
\end{eqnarray}
where $V$ and $ f_{n{\bmk}}$ are the volume of the unit cell and the occupation rate, respectively.

We will derive an explicit expression for the conductivity $\sigma_{\alpha\beta}(t,t')$
defined by Eq.~(\ref{def_conductivity}) in this model.
We express the vector potential for the pump pulse as $\bmA^\rmP(t)$ and for the probe pulse
as $\bmA^\rmp(t)$. They are related to the electric field by $\bmE^{\rmP,\rmp}(t) = -(1/c)(d/dt) \bmA^{\rmP,\rmp}(t)$.
The Hamiltonian with both pump and probe fields is given by, 
\begin{eqnarray}
    && H^{\rmP+\rmp}_{\bmk}(t) \nonumber \\
    &&=\frac{1}{2} \left[{\bmp} +{\bmk} +\frac{1}{c}\left\{\bmA^\rmP(t)+\bmA^\rmp(t)\right\} \right]^2+V({\bmr}) \nonumber\\
    && =H_{\bmK(t)}+\Delta V(t)+ O((A^{\rmp})^2),
    \label{eq:h_tot}
\end{eqnarray}
where we have defined $\bmK(t) = \bmk + \bmA^\rmP(t)/c$ and the perturbation potential
$\Delta V(t)=\frac{1}{c} \left({\bmp} +\bmK(t) \right)\cdot {\bmA}^{\rmp}(t)$.
We express the current density defined by Eq.~(\ref{eq:I}) as ${\bmI}^{\rmP+\rmp}(t)=\bmI^\rmP(t)+\delta \bmI(t)
%+ O((A^{\rm p})^2)
$, where $\delta \bmI(t)$ is the current density defined by Eq.~(\ref{delta_current}).
Using the standard procedure of the time-dependent perturbation theory,  we obtain a formula for the conductivity
defined by Eq.~(\ref{def_conductivity}) as,
\begin{eqnarray}
    \sigma_{\alpha\beta}(t,t')&=&n_e\delta_{\alpha\beta}\theta(t-t')+\frac{2}{V}\theta(t-t') \,
     {\rm Im}\int_{t'}^t dt''  \nonumber\\
    &&\times \sum_{{\bmk}}\sum_{n\neq n'} f_{n{\bmk}}
    [ {\bmP}^{\bmk}_{nn'}(t) ]_{\alpha}[{\bmP}^{\bmk}_{n'n}(t'')]_{\beta},
    \label{eq:sigma_tt}
\end{eqnarray}
where we have introduced the averaged electron number density, $n_e=\sum_{n{\bmk}}f_{n{\bmk}}/V$, and 
the matrix element of the momentum operator,
\begin{equation}
    {\bmP}^{\bmk}_{nn'}(t)= \int_{\rm cell} d\bmr \, w^*_{n{\bmk}}(\bmr,t) \, {\bmp}\, w_{n'{\bmk}}(\bmr,t).
    \label{eq:P}
\end{equation}
Here $w_{n{\bmk}}(\bmr,t)$ is the Bloch orbital in the presence of the pump field only,
which satisfies 
\begin{equation}
i \frac{\partial}{\partial t}  w_{n{\bmk}}(\bmr,t) = H_{\bmK(t)}w_{n{\bmk}}(\bmr,t).
\end{equation}

To proceed further, we approximate the time-dependent Bloch orbital 
$w_{n{\bmk}}({\bmr},t)$ using the Houston function \cite{Yacoby1968,Houston1940},
\begin{equation}
    u^{\rmH}_{n{\bmk}}({\bmr},t)=u_{n\bmK(t)}({\bmr})\,{\exp}\left[-i\int^t_{-\infty}d\tau\,\varepsilon_{n{\bmK}(\tau)}\right].
    \label{eq:houston}
\end{equation}
Then, Eq.~(\ref{eq:P}) is given by
\begin{equation}
    \bmP^\bmk_{nn'}(t) \simeq (\bmp)^{\bmK(t)}_{nn'} \, {\exp} \left[ i\int^t_{-\infty}d\tau\, \omega^{{\bmK}(\tau)}_{nn'} \right],
    \label{eq:P_houston}
\end{equation}
where we have defined $(\bmp)^\bmk_{nn'} = \int_{\rm cell} d\bmr \, u^*_{n\bmk}\, \bmp \, u_{n' \bmk}$ 
and $\omega^{{\bmk}}_{nn'} = \varepsilon_{n{\bmk}}-\varepsilon_{n'{\bmk}}$.
Using this matrix element, we have \cite{Otobe2016}
\begin{eqnarray}
   && \sigma_{\alpha\beta}(t,t')=
   \nonumber\\
   &&   n_e\delta_{\alpha\beta}\theta(t-t')+\frac{2}{V}\theta(t-t') \int_{t'}^t dt'' 
     \sum_{{\bmk}}\sum_{n\neq n'} f_{n\bmk}   \nonumber\\ && \times
     {\rm Im}\, \left[  (p_{\alpha})^{{\bmK}(t)}_{nn'}
    (p_{\beta})^{{\bmK}(t'')}_{n'n}  
    {\exp}\left( i\int_{t''}^t d\tau \, \omega^{{\bmK}(\tau)}_{nn'} \right) \right].
    \label{eq:DFK}
\end{eqnarray}
Substituting Eq.~(\ref{eq:DFK}) into Eq.~(\ref{def_impulsive}), we have
the following expression for the impulsive time-resolved conductivity,
\begin{eqnarray}
   && \tilde{\sigma}^\rmI_{\alpha \beta}(T,\omega) = \frac{i n_e}{\omega} \delta_{\alpha\beta} 
   + \frac{2}{V} \sum_{\bmk}\sum_{n\neq n'}f_{n{\bmk}} \int_0^{\infty} ds\, e^{i\omega s} 
   \int^s_0 dt''   \nonumber\\ 
   &&  \times {\rm Im}\, 
    (p_{\alpha})^{{\bmK}(T+s)}_{nn'}
    (p_{\beta})^{{\bmK}(T+t'')}_{n'n} 
    {\exp}\left( i\int_{t''}^s d\tau\,  \omega^{{\bmK}(T+\tau)}_{nn'} \right) . \nonumber \\
\end{eqnarray}

Below, we will investigate the symmetry properties of the transient conductivity using this expression.
For this purpose, we introduce a quantity $F_{\alpha \beta}^{\omega}[\bmK(t)]$ that is regarded as a functional
of $\bmK(t) = \bmk + \bmA^\rmP(t)/c$,
\begin{eqnarray}
  &&  F_{\alpha \beta}^{\omega}[\bmK(t)]=\sum_{n\neq n'}f_{n{\bmk}} \int_0^{\infty} ds\,
    e^{i\omega s} \int^s_0 dt''
    \nonumber \\ && \times
    (p_{\alpha})^{{\bmK}(s)}_{nn'} (p_{\beta})^{{\bmK}(t'')}_{n'n} 
    {\exp}\left[ i\int_{t''}^s d\tau\, \omega^{{\bmK}(\tau)}_{nn'} \right].
    \label{eq:F_houston}
\end{eqnarray}
Using this quantity, we can rewrite the impulsive time-resolved conductivity as
\begin{eqnarray}
     \tilde{\sigma}^\rmI_{\alpha \beta} (T,\omega) &=& \frac{i n_e}{\omega} \delta_{\alpha\beta} - \frac{i}{V} \sum_{\bmk} 
    \Big(  F_{\alpha \beta}^{\omega}[\bmk+\bmA^\rmP_T(t)/c] 
    \nonumber \\ && 
    - F_{\alpha \beta}^{-\omega}[\bmk+\bmA^\rmP_T(t)/c]^{\ast} \Big),
    \label{eq:sigma_functional}
\end{eqnarray}
where we have introduced a time-shifted vector potential, ${\bmA}^\rmP_T(t)\equiv{\bmA}^\rmP(T+t)$.
Hereafter, we will often use a notation $\tilde{\sigma}_{\alpha \beta}^{\omega}[{\bmA}^\rmP_T(t)]$ 
instead of $\tilde{\sigma}^\rmI_{\alpha \beta} (T,\omega)$ to stress that this conductivity is the functional of ${\bmA}^\rmP_T(t)$.

\subsection{Symmetry properties of the time-resolved conductivity}

Let us consider a point group symmetry operation of the crystal.
We consider a $3\times 3$ orthogonal matrix corresponding to the symmetry operation,  $S=(S_{\alpha \beta})$, 
such that the crystalline potential satisfies $V(S{\bmr})=V({\bmr})$. 
For this symmetry operation, we have 
\begin{eqnarray}
    (p_{\alpha})^{S{\bmk}}_{nn'} &=&\sum_{\beta}S_{\alpha \beta}(p_{\beta})^{{\bmk}}_{nn'},
    \label{eq:symm_P} \\
    \varepsilon_{n,S{\bmk}}&=&\varepsilon_{n{\bmk}},
\end{eqnarray}
where we have used the well-known relation for Bloch orbital, $u_{n,S{\bmk}}({\bmr})=u_{n{\bmk}}(S^{-1}{\bmr})$\cite{}.
Therefore, the functional $F^{\omega}=(F_{\alpha \beta}^{\omega})$ satisfies the following relation 
\begin{equation}
    F^{\omega}[S{\bmK}]=S F^{\omega}[{\bmK}] S^{T},
\end{equation}
where we have used the matrix notation for tensors of rank 2.
Summing up this relation over the Brillouin zone, we get
\begin{eqnarray}
    \sum_{\bmk}S F^{\omega}[{\bmk}+{\bmA}^\rmP_T/c] S^{T}&=&\sum_{\bmk} F^{\omega}[S({\bmk}+{\bmA}^\rmP_T/c)]\nonumber\\
    &=&\sum_{\bmk} F^{\omega}[{\bmk}+S{\bmA}^\rmP_T/c], \nonumber \\
\end{eqnarray}
where we have used the symmetry of the the Brillouin zone: 
$\sum_{\bmk}(\cdots)_{S\bmk}=\sum_{S^{-1}\bmk}(\cdots)_{\bmk}=\sum_{\bmk}(\cdots)_{\bmk}$.
Finally, we obtain the symmetry relation of the time-resolved conductivity as follows:
\begin{equation}
    S\tilde{\sigma}^{\omega}[{\bmA}^\rmP_T]S^{T}=\tilde{\sigma}^{\omega}[S{\bmA}^\rmP_T].
    \label{eq:sym_sigma}
\end{equation}
This is the central result of this paper.

Let us consider a periodic pump field which satisfies the following relation for certain sets of symmetry operation $S$ and period $T_S$,
\begin{equation}
{\bmA}^\rmP(t+T_S)=S{\bmA}^\rmP(t).
\end{equation}
Combining it with Eq.~(\ref{eq:sym_sigma}), we obtain dynamical symmetry properties of the time-resolved conductivity, 
\begin{equation}
 S\tilde{\sigma}^\rmI (T,\omega)S^T = \tilde{\sigma}^\rmI (T+T_S,\omega).
\end{equation}
This relation is a direct consequence of the dynamical symmetry of the Hamiltonian,
\begin{equation}
H_{{\bmk} + \frac{1}{c} {\bmA}^\rmP(t+T_S)} = U^{}_S H_{S^T{\bmk} +\frac{1}{c} {\bmA}^\rmP (t) } U^{\dagger}_S,
\label{eq:dynamical_symmetry_H}
\end{equation}
where $U_S$ is the unitary operator representing $S$. With this operator, the position and the momentum operators satisfy 
$U^{\dagger}_S {\bmr} U^{}_S = S{\bmr}$ and $U^{\dagger}_S {\bmp} U^{}_S = S{\bmp}$, respectively.

Here we mention a few special cases.
If the pump field does not exist, Eq.~(\ref{eq:sym_sigma}) is equal to the usual transformation law for rank 2 tensors. 
If we set the pump field as a static electric field, ${\bmA}^{\rmP}(t)=-c{\bmE}t + ({\rm const.})$, we get 
\begin{equation}
 S\tilde{\sigma} [\bmE] (\omega)S^T = \tilde{\sigma} [S{\bmE}] (\omega),
\end{equation}
where we express the conductivity in the presence of a static electric field $\bmE$ as $\tilde \sigma_{\alpha\beta}[\bmE](\omega)$.
This is a transformation law for the static FKE.

We note that the use of the Houston function [Eq.~(\ref{eq:P_houston})] is not the only route to derive Eq.~(\ref{eq:sym_sigma}).
In the appendix, we discuss the perturbation expansion of the time-resolved conductivity with respect to the pump field to reach 
Eq.~(\ref{eq:sym_sigma}) without the Houston function approximation.

In the following, we investigate the dynamical symmetries of the time-resolved conductivity for each case of crystalline symmetry.

\subsubsection{Dielectrics with inversion symmetry}

We first consider dielectrics with inversion symmetry,
\begin{equation}
    S={\rm diag}(-1,-1,-1),
\end{equation}
exposed to a periodic electric field ${\bmA}^\rmP(t)={\rm Re}\,{\bmA}_0 \, e^{-i\Omega t}$.
From Eq.~(\ref{eq:sym_sigma}), the time-resolved conductivity satisfies the following relation,
\begin{equation}
    \tilde{\sigma}_{\alpha \beta}^{\omega}[{\bmA}^\rmP_T]=\tilde{\sigma}_{\alpha \beta}^{\omega}[-{\bmA}^\rmP_T].
\end{equation}
Namely, we get
\begin{equation}
    \tilde{\sigma}^\rmI_{\alpha \beta} (T,\omega)=\tilde{\sigma}^\rmI_{\alpha \beta} \left(T+  \frac{T_\Omega}{2}, \omega \right).
    \label{eq:inversion}
\end{equation}
Therefore the transient conductivity $\tilde{\sigma}_{\alpha \beta}^\rmI (T,\omega)$ shows an oscillatory behavior with an even multiple 
of the frequency of the field ($2\Omega$ oscillation).
%That is, $\tilde{\sigma}_{\alpha \beta}^{(n)}(\omega)$ is zero if $n$ is odd.
This is consistent with previous works reporting the $2\Omega$ oscillation in the transient optical response for inversion symmetric dielectrics \cite{Otobe2016,Lucchini2016}.
%This result can also be understood as the well-known property of the nonlinear responses in an inversion symmetric crystal, whose even-order responses are forbidden by the inversion symmetry.
%Here we note that the second order nonlinear response corresponds to the $\Omega$ oscillation of the time-resolved conductivity (see Appendix). 

\subsubsection{Dielectrics with reflection symmetry under a linearly polarized field}

We next consider dielectrics with reflection symmetry.
We consider a system that has reflection symmetry with respect to the $xy$ plane,
\begin{equation}
    S={\rm diag}(1,1,-1),
    \label{eq:S_ref}
\end{equation}
and a linearly-polarized periodic field along the $z$ direction:
\begin{equation}
    {\bmA}^\rmP(t) = (0,0,A_0\, {\cos} \Omega t).
\end{equation}

Similarly to Eq.~(\ref{eq:inversion}), for $\alpha,\beta=x,y$ or $\alpha=\beta=z$ components, we obtain
\begin{equation}
    \tilde{\sigma}^\rmI_{\alpha \beta} (T,\omega)=\tilde{\sigma}^\rmI_{\alpha \beta} \left( T+\frac{T_{\Omega}}{2},\omega \right),
    \label{eq:2omega_reflection}
\end{equation}
and therefore these components have a feature of the $2\Omega$ oscillation.
%Namely, $\tilde{\sigma}_{\alpha \beta}^{(n)}(\omega)$ survives only for even $n$.

For $\alpha=x,y$ and $\beta=z$ (or $\alpha=z$ and $\beta=x,y$), we obtain
\begin{equation}
    \tilde{\sigma}^\rmI_{\alpha \beta} (T,\omega)=-\tilde{\sigma}^\rmI_{\alpha \beta} \left( T+\frac{T_{\Omega}}{2},\omega \right).
\end{equation}
These components show oscillation with a period of $\Omega$.
Namely, the transient absorption shows no more symmetry than that of the applied field. 
%Now there is no restriction for  $\tilde{\sigma}_{\alpha \beta}^{(n)}(\omega)$.

In Ref.~\onlinecite{Sato2018}, it has been reported that there appears $2\Omega$ oscillation originating from 
the reflection symmetry and $\Omega$ oscillation for a system without any symmetry.

\subsubsection{Dielectrics with $N$-fold rotational symmetry under a circularly-polarized field}

We consider dielectrics with $N$-fold rotational symmetry around the $z$ axis and a circularly-polarized periodic field,
\begin{equation}
    {\bmA}^\rmP(t) = (A_0\, {\cos} \Omega t, A_0\, {\sin} \Omega t,0).
    \label{eq:A_circular}
\end{equation}

In this case, it is convenient to use a complex-valued function in the $xy$ plane,
\begin{equation}
    k_c + A_T(t)/c,\quad A_T(t)=A_0\,e^{i\Omega (T+t)},
\end{equation}
where $k_c$ is given by $k_c= k_x + i k_y$.
Using the  $N$-fold  rotation operation around the $z$ axis,
\begin{equation}
    S:\, k_c \longrightarrow k_c \,e^{2\pi i/N},
    \label{eq:S_Nrot}
\end{equation}
and Eq.~(\ref{eq:sym_sigma}), we obtain 
\begin{eqnarray}
    \tilde{\sigma}_{zz}^{\omega}[A_T]&=&\tilde{\sigma}_{zz}^{\omega}[A_T\,e^{2\pi i/N}]\nonumber\\
    &=&\tilde{\sigma}_{zz}^{\omega}[A_{T+\frac{1}{N}T_{\Omega}}].
\end{eqnarray}
Namely, we obtain
\begin{equation}
    \tilde{\sigma}^\rmI_{zz} (T,\omega)=\tilde{\sigma}^\rmI_{zz} \left(T+\frac{T_{\Omega}}{N},\omega \right),
    \label{eq:circular_Nomega}
\end{equation}
and therefore this component has the $N\Omega$ oscillation.
%The Fourier component $\tilde{\sigma}_{\alpha \beta}^{(n)}(\omega)$ survives only for $n$ which is a multiple of $N$.

\subsubsection{Dielectrics with $N$-fold improper rotational symmetry under a circularly-polarized field}

For the case of dielectrics with an improper rotational symmetry, we may consider a product of Eq.~(\ref{eq:S_ref}) and Eq.~(\ref{eq:S_Nrot}).
Since the reflection Eq.~(\ref{eq:S_ref})  does not change Eq.~(\ref{eq:A_circular}), the same discussion as given above holds.
Therefore we get Eq.~(\ref{eq:circular_Nomega}) again.

\section{First-principles pump-probe calculations \label{sec:first-principles}}

\subsection{Formalism}

In order to verify the symmetry properties of the transient absorption spectroscopy, we perform real-time TDDFT calculations.
Since we have made several assumptions and approximations in developing the analytical consideration,
numerical calculations will be useful to confirm their validity.

The details of the computational methods have already been reported elsewhere\cite{Otobe2016,Yabana1996,Bertsch2000,Otobe2008}.
We solve TDKS equation for the pump-probe process,
\begin{eqnarray}
    i\frac{\partial}{\partial t}v_{n{\bmk}}^{\rm KS}({\bmr},t)= \Bigg[
    \frac{1}{2} \left[{\bmp} +{\bmk} +\frac{1}{c}\left\{{\bmA}^\rmP(t)+{\bmA}^{\rmp}(t)\right\} \right]^2
    \nonumber \\
    +V_{\rm H}({\bmr},t)+V_{\rm xc}({\bmr},t)+V_{\rm ion}({\bmr})
   \Bigg] v_{n{\bmk}}^{\rm KS}({\bmr},t), \quad\quad
    \label{eq:tdks}
\end{eqnarray}
where $V_{\rm H}({\bmr},t)$, $V_{\rm xc}({\bmr},t)$, and $V_{\rm ion}({\bmr})$ are the electron-electron Hartree, 
the exchange-correlation, and the electron-ion potential, respectively.
If we ignore the time dependence of $V_{\rm H}({\bmr},t)$ and  $V_{\rm xc}({\bmr},t)$, the Hamiltonian of Eq.~(\ref{eq:tdks}) 
coincides with Eq.~(\ref{eq:h_tot}).
In most dielectric materials, differences of the Hartree and the exchange-correlation potentials from those in the ground state 
are small and are not significant \cite{Lucchini2016,Tancogne-Dejean2017}.
We employ the norm-conserving pseudopotential \cite{Troullier1991} for $V_{\rm ion}({\bmr})$ and the adiabatic 
local-density approximation \cite{Perdew1981} for $V_{\rm xc}({\bmr},t)$.
For simplicity of implementation, we ignore the exchange-correlation term of the vector potential,  
${\bmA}_{\rm xc}(t)$, which should be included for a rigorous treatment of the exchange-correlation effects 
of infinite periodic systems \cite{Vignale1996}.

For the time profile of the pump and the probe pulses, we employ the following form:
\begin{eqnarray}
{\bmA}^\rmP(t)&=&-\frac{c {\bmE}^\rmP_0}{\Omega} \cos^2 \frac{\pi t}{T_{\rmP}} \sin \Omega t , 
\nonumber \\ && \qquad (-T_{\rmP}/2 < t < T_{\rmP}/2), %!!!!!
\\
{\bmA}^{\rmp}(t)&=&-\frac{c {\bmE}^{\rmp}_0}{\omega_{\rmp}} \cos^2 \frac{\pi (t-T)}{T_{\rmp}} \sin  \left[ \omega_{\rmp} (t-T) \right], 
\nonumber \\ && \qquad (-T_{\rmp}/2 < t-T < T_{\rmp}/2), %!!!!!
\end{eqnarray}
where $T_{\rmP}$ and $T_{\rmp}$ are the full duration of the pump and probe pulses, respectively.
Here $ \omega_{\rmp}$ stands for the average frequency of the probe pulse.
Again, $T$ is the central time of the probe pulse and now corresponds to the delay time between the pump and probe pulses.
In the calculations below, we will use a sufficiently large value for $T_{\rmP}$ so that the pump field ${\bmA}^\rmP(t)$ can be 
regarded as periodic as given in Eq.~(\ref{eq:periodicity_A}).

Using the TDKS orbital $v_{n{\bmk}}^{\rm KS}({\bmr},t)$ instead of $v_{n{\bmk}}({\bmr},t)$ in Eq.~(\ref{eq:I}), 
we can obtain the spatially averaged current density corresponding to the pump-probe process.
We include a term originating from the nonlocality of the pseudopotential as well \cite{Bertsch2000}.
We calculate the energy transfer, or work, from the probe pulse to electron in the unit cell of dielectrics,
\begin{equation}
\delta W = \int dt\, \delta \bmI(t) \cdot \bmE^\rmp(t),
\label{eq:energy_transfer_probe}
\end{equation}
where $\delta \bmI(t)$ is defined by Eq.~(\ref{delta_current}).
If we make a frequency-resolved analysis, we obtain $A(\omega)$ defined in Eq.~(\ref{Aw}).
To confirm the symmetry property, however, it is sufficient to examine without frequency resolution.

\subsection{Results}

\begin{figure}
    \includegraphics[keepaspectratio,width=\columnwidth]{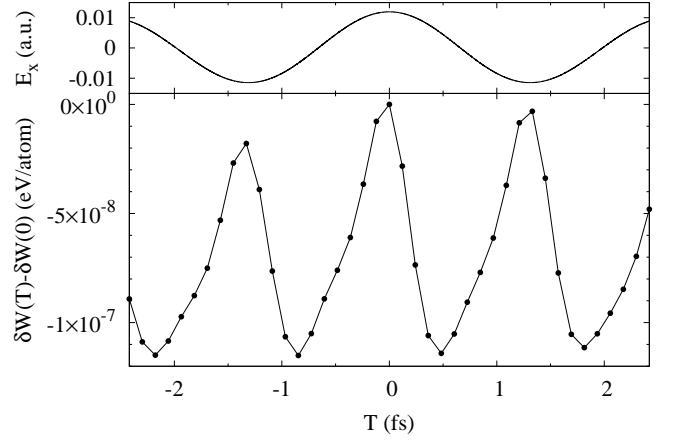}
    \caption{\label{fig:x} 
  Energy transfer [Eq.~(\ref{eq:energy_transfer_probe})] as a function of the delay time $T$ with the pump and the probe pulses of the  $x$   polarization (lower panel).
  Upper panel: time-profile of pump electric field for comparison.
    }
\end{figure}

\begin{figure}
    \includegraphics[keepaspectratio,width=\columnwidth]{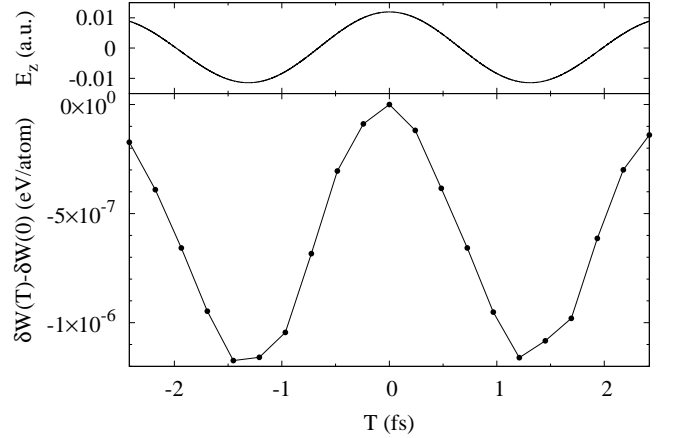}
    \caption{\label{fig:z} 
   The same as in Fig.~\ref{fig:x}, but for the $z$  polarization.
    }
\end{figure}

\begin{figure}
    \includegraphics[keepaspectratio,width=\columnwidth]{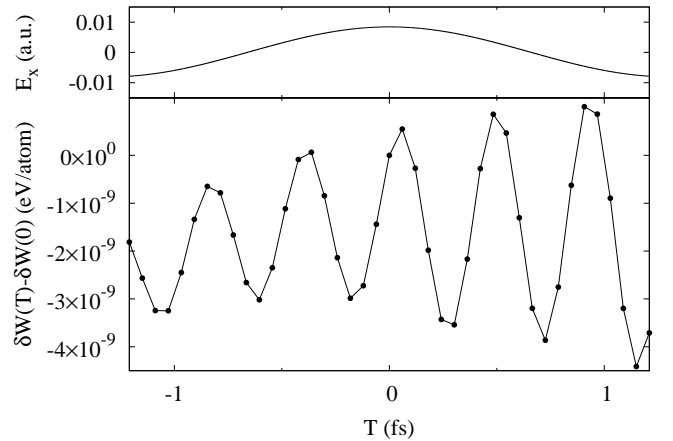}
    \caption{\label{fig:circular} 
   The same as in Fig.~\ref{fig:x}, but for the circular polarized pump field in the $xy$ plane and the $z$ polarized probe field.
    }
\end{figure}

We present results for 4H-SiC crystal, which has $C_{6v}$ point group symmetry.
We employ the pump pulse of $\hbar \Omega= 1.55$ eV and the probe pulse of $\hbar \omega_{\rmp}= 40$ eV.
We set $T_{\rmP}$ and $T_{\rmp}$ to 20 fs and 1 fs, respectively.
The field strength of the pump and probe pulses are set to $E^{\rmP}_0=6.1\times 10^{-1}$ V/{\AA} and $E_0^{\rmp}=2.7\times 10^{-2}$ V/{\AA}, respectively, 
where the latter value is small enough to justify the perturbative treatment for the probe process. 

We performed the calculations by using the open-source software SALMON (Scalable Ab-initio Light-Matter simulator for
Optics and Nanoscience) \cite{Noda2019} which has been developed in our group and is available
from the website, Ref.~\onlinecite{SALMON_web}.
The code solves Eq.~(\ref{eq:tdks}) in the time domain with the real-space finite-difference method in the 3D Cartesian coordinate.
We employ a real-space grid of $20\times 32\times 64$ for the rectangular unit cell of 16 atoms and a k-space grid of $20\times 12\times 6$ for the Brillouin zone sampling.
The Taylor expansion method is used for the time evolution with a time step of $dt = 0.02$ in atomic units. 
The number of time steps is typically 42 000. 

Figure~\ref{fig:x} shows the energy transfer as a function of the delay time $T$ using the pump and probe pulses of the linear polarization along the $x$ axis.
The energy transfer shows an oscillation with the frequency of twice the pump frequency $\Omega$.
This $2\Omega$ oscillation is due to the reflection symmetry of the hexagonal structure with respect to the $x$ axis 
[see Eq.~(\ref{eq:2omega_reflection})].

Figure~\ref{fig:z} shows the energy transfer for the pump and probe pulses with the $z$ polarization.
Since the hexagonal crystal structure has no symmetry along the $z$ axis, the energy transfer 
indicates the $\Omega$ oscillation.

Figure~\ref{fig:circular} shows the energy transfer for the circularly-polarized pump field in the $xy$ plane and the linearly-polarized probe field in the $z$ direction.
The energy transfer shows the $6\Omega$ oscillation. This is expected from Eq.~(\ref{eq:circular_Nomega}) and the 6-fold rotational symmetry of 
the crystal structure.

We thus find that the oscillatory structures of the energy transfer seen in our calculations are all consistent with 
our analytical investigation for three cases of different symmetry properties.
The difference in the magnitude of the modulation seen in Fig.~\ref{fig:x}-\ref{fig:circular} can be understood from perturbation orders 
with respect to the pump field (see Appendix).
Eq.~(\ref{eq:sigma_perturbation}) suggests that the $n\Omega$ oscillation of the time-resolved conductivity starts to appear in the $(n+1)$th order nonlinear response.
Therefore, the magnitude of the $6\Omega$ oscillation (Fig.~\ref{fig:circular}) is 5-order smaller than that of the $\Omega$ oscillation (Fig.~\ref{fig:z}), and so on.

In Ref.~\onlinecite{Otobe2016-2}, it was reported that an oscillation of the DFKE response in diamond crystal almost disappears for circularly polarized light
from TDDFT calculation. It could be understood from the order of the perturbation: $4\Omega$ oscillation signal is expected from the 4-fold rotational symmetry 
of the diamond structure. However, the magnitude is 2-order smaller than the $2\Omega$ oscillation signal seen in the linearly-polarized probe field.
Because of the difference of two-orders in the perturbation, the signal in the circularly-polarized pulse looks extremely small.

\section{Conclusion \label{sec:conclusion}}

We have discussed the relation between crystalline symmetry and the signal of attosecond transient absorption spectroscopy
caused by the dynamical Franz-Keldysh effect.
We found that an oscillation property of a transient conductivity in a laser-exposed crystal is determined by a transformation low 
of the transient conductivity tensor under a crystalline symmetry operation.
We obtained selection rules for the frequency of the laser-driven oscillation of the probe response in the pump-probe time domain.
First-principles calculations based on the time-dependent density functional theory confirmed validity of these selection rules for 
several crystalline symmetries. 
This work paves the way for understanding the physics of attosecond transient absorption spectroscopy in crystalline solids.

\begin{acknowledgements}
This research was supported by JST-CREST under grant number JP-MJCR16N5, and by MEXT as a priority issue theme 7 to be tackled by using Post-K Computer, and by JSPS KAKENHI Grant Numbers 15H03674. Calculations are carried out at Oakforest-PACS at JCAHPC under the support by Multidisciplinary Cooperative Research Program in CCS, University of Tsukuba.
\end{acknowledgements}

\appendix

\section{Perturbation expansion}

In this Appendix, we develop a perturbation theory for the modulation of the conductivity
in which the perturbation expansion is carried out with respect to the pump field of the form,
${\bf A}^{\rmP}(t)=\sum_p \tilde{\bf A}(\Omega_p) e^{-i\Omega_p t}$.
The Hamiltonian of Eq.~(\ref{eq:tdse}) can be rewritten as
\begin{equation}
	H_{{\bf K}(t)}=H_{\bf k} + \Delta V_{\rmP}(t) + c_{\bf k}(t),
\end{equation}
where $\Delta V_{\rmP}(t) = {\bf p}\cdot {\bf A}^{\rmP}(t)/c $ and $c_{\bf k}(t)$ are the perturbation potential and the remaining c-number term, respectively.
The time-dependent Bloch orbital $w_{n{\bf k}}({\bf r},t)$ is expanded as
\begin{equation}
	|w_{n{\bf k}}(t)\rangle =\sum^{\infty}_{N=0} |w^{(N)}_{n{\bf k}}(t)\rangle,    
\end{equation}
where the subscript ``$(N)$" stands for the $N$th perturbation order.
$N$th order orbital is expressed as
\begin{eqnarray}
	|w^{(N)}_{n{\bf k}}(t)\rangle &=&
    e^{-i \varepsilon_{n{\bf k}} t}\sum_m |u_{m{\bf k}} \rangle \sum_{p_1 \cdots p_N} 
    \nonumber \\
     &\times &
    \tilde{C}^{(N)}_{mn{\bf k}}(\Omega_{p_1},\cdots, \Omega_{p_N})
    e^{  -i \sum^N_{i=1} \Omega_{p_i} t }.
    \label{eq:w_N}
\end{eqnarray}
Here the coefficient $\tilde{C}^{(N)}_{mn{\bf k}}$ is determined from $\tilde{C}^{(0)}_{mn{\bf k}}=\delta_{mn}$ and a recursion formula,
\begin{eqnarray}
	\tilde{C}^{(N)}_{mn{\bf k}}(\Omega_{p_1},\cdots, \Omega_{p_N})= \frac{1}{c}
    \sum_l \frac{({\bf p})^{\bf k}_{ml} \cdot \tilde{\bf A}(\Omega_{p_N})}{ \sum^{N}_{i=1}\Omega_{p_i} - \omega^{\bf k}_{mn}}
    \nonumber \\
     \times
    \tilde{C}^{(N-1)}_{ln{\bf k}}(\Omega_{p_1},\cdots, \Omega_{p_{N-1}}).
    \label{eq:C_N}
\end{eqnarray}
Up to here the procedure is mostly the same as that of the ordinary time-dependent perturbation theory for 
nonlinear optical susceptibility\cite{Boyd}.

Hereafter we will deal with monochromatic light: ${\bf A}^{\rmP}(t)={\rm Re} \, {\bf A}_0\, e^{-i\Omega t}.$
Namely we set the frequency as $\Omega_p = p\Omega, \, (p=\pm)$ and the coefficients as $\tilde{\bf A}(\Omega_{+}) ={\bf A}_0 /2 ,$ $\tilde{\bf A}(\Omega_{-}) ={\bf A}_0^{\ast} /2$. 
Then we rewrite the matrix element of Eq.~(\ref{eq:P}) as
\begin{eqnarray}
	{\bf P}^{\bf k}_{nn'}(t)
    = e^{i\omega^{\bf k}_{nn'}t } \sum_{\nu=-\infty}^{\infty} e^{-i \nu \Omega t} \,\tilde{\bf P}^{\bf k}_{nn'}(\nu),
\end{eqnarray}
where
\begin{eqnarray}
\tilde{\bf P}^{\bf k}_{nn'}(\nu) = \sum_{N, N'=0}^{\infty} 
    \sum_{p_1 \cdots p_N} \sum_{q_1 \cdots q_{N'}} 
    \delta_{\nu, -\sum p_i + \sum q_i} \nonumber \\
   \times \sum_{m m'} 
    \tilde{C}^{(N) \ast}_{mn{\bf k}}(\Omega_{p_1},\cdots, \Omega_{p_N}) 
    \nonumber \\
    \times
    ({\bf p})^{\bf k}_{mm'}  
    \tilde{C}^{(N')}_{m'n'{\bf k}}(\Omega_{q_1},\cdots, \Omega_{q_{N'}}).
\end{eqnarray}
Using Eq.~(\ref{eq:sigma_tt}) and Eq.~(\ref{def_impulsive}), we get
\begin{eqnarray}
\tilde{\sigma}^{\rmI}_{\alpha \beta} (t,\omega) &=&
\frac{i n_e}{\omega} \delta_{\alpha\beta} + \frac{i}{V} \sum_{\bf k}\sum_{n\neq n'}f_{n{\bf k}} 
\sum_{\nu \nu'}  
\nonumber \\
 && \times 
\frac{ \left[ \tilde{\bf P}^{\bf k}_{nn'}(\nu) \right]_{\alpha}
\left[ \tilde{\bf P}^{\bf k}_{n'n}(\nu') \right]_{\beta} \,  e^{-i(\nu+\nu')\Omega t}  }
{ \left\{ \omega^{+} - (\nu +\nu')\Omega \right\} (\omega^{+} - \nu \Omega +\omega^{\bf k}_{nn'}  ) }
\nonumber \\
 &&   - (\omega \rightarrow -\omega)^{\ast} ,
    \label{eq:sigma_perturbation}
\end{eqnarray}
where $\omega^{+} = \omega + i0$.

From Eq.~(\ref{eq:w_N}) and Eq.~(\ref{eq:C_N}), the $t$ dependence of Eq.~(\ref{eq:sigma_perturbation})
can be absorbed in the vector potential of the pump field  by redefinition, ${\bf A}_0 \rightarrow {\bf A}_{t,\,0}\equiv{\bf A}_0 \, e^{-i\Omega t}$.
Here  ${\bf A}_{t,\,0}$ is equal to the coefficient of the time-shifted vector potential ${\bf A}^{\rmP}_t(x)={\bf A}^{\rmP}(t+x)$.
Furthermore, Eq.~(\ref{eq:sym_sigma}) is valid in this situation because the coefficient $\tilde{C}^{(N)}_{mn{\bf k}}$ is unchanged under a transformation $S: {\bf k}\rightarrow S{\bf k},\,{\bf A}^{\rmP}_t\rightarrow S{\bf A}^{\rmP}_t$.
Therefore we can repeat the argument described below Eq.~(\ref{eq:sym_sigma}).
We note that for the case of ${\bf A}^{\rmP}(t)=0$, Eq.~(\ref{eq:sigma_perturbation}) agrees with the well-known Kubo–Greenwood formula of the conductivity since $\nu=\nu'=0$ and  $\tilde{\bf P}^{\bf k}_{nn'}(\nu)=({\bf p})^{\bf k}_{nn'}$.

　　　
%\bibliography{article,misc}

\end{document}